\documentclass[pra,a4paper,aps,superscriptaddress,twocolumn,longbibliography,10pt]{revtex4-2}

\usepackage{amsmath, empheq,braket,amsthm,amssymb,graphics,amsfonts,array,color,subfigure, mathrsfs}
\definecolor{myurlcolor}{rgb}{0,0,0.7}
\definecolor{myurlcolor1}{rgb}{0,0.7,0.1}
\definecolor{myrefcolor}{rgb}{0,0,0.7}
\usepackage[unicode=true,pdfusetitle, bookmarks=true,bookmarksnumbered=false,bookmarksopen=false, breaklinks=false,pdfborder={0 0 0},backref=false,colorlinks=true, linkcolor=myrefcolor,citecolor=myurlcolor1,urlcolor=myurlcolor]{hyperref}

\usepackage{dcolumn}
\usepackage{bm}
\usepackage{bbold}
\usepackage{dsfont} 

\newtheorem{theo}{Theorem}

\newcommand{\kb}[2]{\left| #1 \vphantom{#2} \right>\left< #2 \vphantom{#1} \right|} 
\newcommand{\proj}[1]{\kb{#1}{#1}} 
\newcommand{\tr}{\mathrm{Tr}}
\newcommand{\bs}{\hat{U}^{\mathrm{BS}}_{\eta}}
\newcommand{\bsd}{\hat{U}^{\mathrm{BS} \dagger}_{\eta}}
\newcommand{\dd}{\mathrm{d}}

\begin{document}

\title{A central limit theorem for partially distinguishable bosons}

\author{Marco Robbio}
\email{marco.robbio@ulb.be}
\affiliation{Centre for Quantum Information and Communication, \'Ecole polytechnique de Bruxelles, CP 165/59, Universit\'e libre de Bruxelles, 1050 Brussels, Belgium}
\affiliation{International Iberian Nanotechnology Laboratory (INL), Av. Mestre Jos\'e Veiga, 4715-330 Braga, Portugal}

\author{Michael G. Jabbour}
\email{michael.jabbour@ulb.be}
\affiliation{Centre for Quantum Information and Communication, \'Ecole polytechnique de Bruxelles, CP 165/59, Universit\'e libre de Bruxelles, 1050 Brussels, Belgium}

\author{Leonardo Novo}
\email{leonardo.novo@inl.int}
\affiliation{International Iberian Nanotechnology Laboratory (INL), Av. Mestre Jos\'e Veiga, 4715-330 Braga, Portugal}
\affiliation{Centre for Quantum Information and Communication, \'Ecole polytechnique de Bruxelles, CP 165/59, Universit\'e libre de Bruxelles, 1050 Brussels, Belgium}

\author{Nicolas J. Cerf}
\email{nicolas.cerf@ulb.be}
\affiliation{Centre for Quantum Information and Communication, \'Ecole polytechnique de Bruxelles, CP 165/59, Universit\'e libre de Bruxelles, 1050 Brussels, Belgium}

\begin{abstract}
 The quantum central limit theorem derived by Cushen and Hudson provides the foundations for understanding how subsystems of large bosonic systems evolving unitarily do reach equilibrium. It finds important applications in the context of quantum interferometry, for example, with photons. A practical feature of current photonic experiments, however, is that photons carry their own internal degrees of freedom pertaining to, \textit{e.g.}, the polarization or spatiotemporal mode they occupy, which makes them partially distinguishable. 
 The ensuing deviation from ideal indistinguishability is well known to have observable consequences, for example in relation with boson bunching, but an understanding of its role in bosonic equilibration phenomena is still missing. Here, we generalize the Cushen-Hudson quantum central limit theorem to encompass scenarios with partial distinguishability, implying an asymptotic convergence of the subsystem's reduced state towards a multimode Gaussian state defined over the internal degrees of freedom. While these asymptotic internal states may not be directly accessible, we show that particle number distributions carry important signatures of distinguishability, which may be used to diagnose experimental imperfections in large boson sampling experiments.     
\end{abstract}

\maketitle

\section{Introduction}

\begin{figure}[t!]\label{Fig.1}
\centering
\includegraphics[scale=0.5]{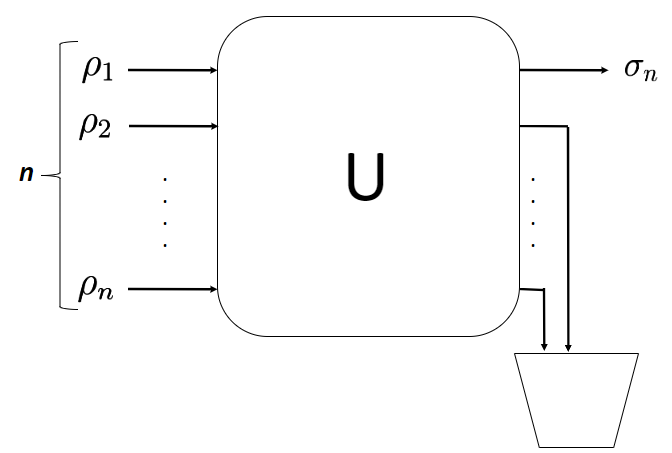}
\caption{The QCLT describes the asymptotic behavior of the reduced output state $\sigma_n$ resulting from the unbiased linear interferometer $U$ acting on $n$ identical single-mode bosonic states $\rho_i\equiv \rho$, each one entering a different spatial mode of the interferometer. In the large-$n$ limit, $\sigma_n$ tends towards a Gaussian state. In this work, we generalize the QCLT to the case where the $\rho_i$'s represent possibly different multimode states associated with internal degrees of freedom (modeling  partially distinguishable bosons, as depicted in Fig.~\ref{fig:V_i}), assuming the unbiased interferometer $U$ only couples the spatial modes but does not affect these internal degrees of freedom.}
\end{figure}

 One of the most fundamental problems in quantum statistical mechanics is to understand how quantum systems reach equilibrium. While the unitary evolution in quantum mechanics is inherently reversible, it is often observed that small subsystems of large quantum systems evolving unitarily are well described by an equilibrium state modelled by a generalized Gibbs ensemble~\cite{Gogolin_2016}. The foundations for the theoretical understanding of such equilibration phenomena are laid out by the so-called quantum central limit theorems (QCLTs)
 \cite{Central, QCLT1,QCLT2, QCLT3,QCLT4,QCLT6,PhysRevLett.96.080502,QCLT5}.  For non-interacting bosonic systems, the seminal work of Cushen and Hudson \cite{Central} explains how a linear interference process between $n$ identical states, characterized by an unbiased sum of the position/momentum coordinates, results in an output state that is Gaussian when $n$ is large~\cite{Weedbrook2012}. In photonic devices, this process can be observed via a balanced interferometer: if identical photonic states are sent in each input port of an ``$n$-splitter", the reduced state of a single output port will converge to a Gaussian state (see Fig.~\ref{Fig.1}). This \emph{Gaussification} of reduced states can also be proved rigorously for other systems of physical interest, such as non-interacting bosons evolving under lattice Hamiltonians \cite{cramer2008lattice, cramer2010QCLT_correlated, gluza2019equilibration}.   

Recent experimental developments in photonic and atomic platforms, motivated by the rise of quantum computation and simulation, allow for an unprecedented control over bosonic systems, providing an ideal testbed to study equilibration phenomena \cite{langen2015experimental, somhorst2023quantum, young2023atomic}. However, an important limitation in photonic systems, in particular, lies in that photons entering an interferometry unavoidably exhibit slight differences in their internal degrees of freedom (corresponding to the polarization, spatial, time or frequency of the mode they occupy), hence making them partially distinguishable. The study of linear interference with partially distinguishable photons has recently been active research topic, not only because distinguishability may limit demonstrations of quantum computational advantage \cite{renema_dist, shi2022part_dist_GBS} via the boson sampling problem~\cite{bosonsampling}, but also due to interesting physical phenomena that can be observed such as collective photonic phases \cite{shchesnovich2018collective, Menssen2017} or anomalous boson bunching effects \cite{Bunching}. However, equilibration phenomena involving partially distinguishable photons are still poorly understood as of today. 

It is, nevertheless, expected that distinguishability plays an important role in the equilibration of subsystems resulting from large linear interference experiments~\cite{somhorst2023quantum}. If we take, as an example, the case of $n$ indistinguishable single photons at the input of the unbiased interferometer of Fig.~\ref{Fig.1}, it is known from Cushen and Hudson's theorem that the asymptotic state of a single output mode is given by a thermal state. The corresponding photon number distribution is thus a geometric distribution with average photon number $1$. In contrast, the same experiment involving fully distinguishable photons, leads to a Poisson distribution with the same average photon number. This is because in the latter case, the process is governed by purely classical statistics (i.e., the photons are scattered as classical balls). Since each photon has a probability $1/n$ of leaving the interferometer in a single output mode, say the first one, the photon number distribution is given by the convolution of $n$ Bernoulli random variables.

In this work, we formulate a bosonic QCLT taking explicitly into account partial distinguishability. This allows us  not only to recover in the extreme cases the behavior of indistinguishable and distinguishable particles, but also to understand equilibration in more complex partial distinguishability scenarios. To do so, we describe the asymptotic Gaussian state obtained in a single output port of an unbiased interferometer, taking into account the standard assumption that the interferometer couples the $n$ spatial modes but leaves the internal degrees of freedom invariant \cite{Tichy2015}. For interferometers with $n$ single photons -- possibly in distinct internal states --  at the input, our results predict that this asymptotic state is given by a generalized Gibbs ensemble, whose different ``temperatures" are given by the eigenvalues of the distinguishability matrix, \textit{i.e.}, the Gram matrix containing all relational information on the photonic internal states.  We believe our results will help understanding signatures of partial distinguishability in equilibration phenomena involving non-interacting bosons, which could be of immediate interest for the validation of large-scale photonic and atomic boson sampling experiments \cite{Wang2019, zhong2020quantum, madsen2022quantum, young2023atomic}.

The paper is structured as follows. In Section~\ref{sec:Preliminaries}, we provide a brief overview of relevant techniques used in this work to treat continuous-variable systems as well as the linear interference of partially distinguishable bosons. In Section~\ref{sec:A quantum central limit theorem with partial distinguishability}, we present our main result -- a QCLT for partially distinguishable bosons as well as its interpretation in particular situations of physical interest. In Section~\ref{sec:GF}, we focus on the derivation of photon number distributions in specific partial distinguishability scenarios which interpolate between distinguishable and indistinguishable bosons. Finally, we present our conclusions in Section~\ref{sec:conclusions}. 

%
%
\section{Preliminaries}\label{sec:Preliminaries}
\subsection{Bosonic systems and Gaussian states}
Our work focuses on bosonic systems, whose paradigm can be found in $n$ radiation modes of the electromagnetic field~\cite{Weedbrook2012}. Each bosonic mode can be described in an infinite-dimensional Hilbert space and is associated with a pair of bosonic field operators $a_i$ and $a_i^\dagger$, where the index $i=1, \ldots, n$ labels the mode. These so-called annihilation and creation operators satisfy the bosonic commutation relations $[a_i,a_j^\dagger] =  \frac{1}{2}\, \delta_{ij} \, \mathds{1}$, $[a_i,a_j] = 0$, $[a_i^\dagger,a_j^\dagger] = 0$, where $\mathds{1}$ is the infinite-dimensional identity operator (note that throughout, we take the convention $\hbar=1$) . Alternatively, the system can be described using the quadrature operators $\{ \hat{x}_j, \hat{p}_j \}_{j=1}^n$ defined as (with $i$ being the imaginary unit in the equation below)
\begin{equation}
    \hat{x}_j = \frac{a_j+ a_j^{\dagger}}{\sqrt{2}}, \quad \hat{p}_j =\frac{ i(a_j^{\dagger} - a_j)}{\sqrt{2}}.
\end{equation}
If these operators are arranged in the vector $\hat{\boldsymbol{r}} = (\hat{x}_1, \hat{p}_1, \ldots, \hat{x}_n, \hat{p}_n)$, they satisfy the relation
\begin{equation}
    [\hat{r}_k, \hat{r}_l] =  i \Omega_{kl}, \quad
    \boldsymbol{\Omega} = \bigoplus_{k=1}^N \begin{pmatrix} 0 & 1 \\ -1 & 0 \end{pmatrix}.
\end{equation}
Each of the $n$ bosonic modes is modelled by a quantum harmonic oscillator, so that the free Hamiltonian of each mode is given by $\hat{H}_i = \hat{N}_i $, where $\hat{N}_i = a_i^\dagger a_i$ denotes the number operator. The latter can be decomposed in the infinite-dimensional Fock basis $\{ \ket{j} \}_{j \in \mathbb{N}}$ as $\hat{N}_i = \sum_{j=0}^{\infty} j \proj{j}$, where we have chosen to omit the label $i$ of the mode on the states $\ket{j}$ for notational simplicity.

A state of the bosonic system is described by a density matrix $\rho$ acting on the tensor product of the $n$ infinite-dimensional Hilbert spaces, but it can be equivalently characterized by its so-called normally-ordered characteristic function
\begin{equation}
    \chi_{\rho}(\boldsymbol{z}) = \tr(D\left(\boldsymbol{z})\, \rho\right),
\end{equation}
where $D(\boldsymbol{z}) = e^{\boldsymbol{z} \cdot \boldsymbol{a}^\dagger} e^{- \boldsymbol{z}^* \cdot \boldsymbol{a}}$ are displacement operators, with $\boldsymbol{z} \in \mathbb{C}^n$, $\boldsymbol{a}^\dagger = (a_1^{\dagger}, \ldots, a_n^{\dagger})$, and $\boldsymbol{a} = (a_1, \ldots, a_n)$.
The choice of the normal-ordered displacement operator and, consequently, the normal-ordered characteristic function is particularly useful for deriving the particle number distribution, which can be obtained from the following relation~\cite{Displacement}:
\begin{equation}\label{probabilities}
p_{m}=\langle m|\rho|m\rangle=\int \frac{\dd^{2n} \boldsymbol{z}}{\pi^{n}} e^{-|\boldsymbol{z}|^{2}} \langle m|D(\!-\!\boldsymbol{z})|m\rangle \, \chi_{\rho}(\boldsymbol{z}).
\end{equation}
Notably, Eq.~\eqref{probabilities} depends on the values $\langle m|D(-\boldsymbol{z})|m\rangle=\mathit{L}_{m}(|\boldsymbol{z}|^{2})$, where $\mathit{L}_{m}$ is the $m$-th order Laguerre polynomial. Moreover, choosing the normal-ordered characteristic function also simplifies the computation of the expectation values of the creation and annihilation operators, which can be computed through the following relation:
\begin{equation}
\langle (a_{l}^{\dagger})^{m}(a_{k})^{n}\rangle=\frac{\partial^{m+n}}{\partial (z_{l})^{m}\partial (-z_{k}^{*})^{n}} \chi_{\rho}(\boldsymbol{z}){\bigg |}_{\boldsymbol{z}=0}.
\end{equation}
Quantum states of $n$ modes can furthermore equivalently be characterized by their so-called Wigner quasidistributions $W_\rho \in \mathbb{R}$, which are related to their characteristic functions through a simple Fourier transform, i.e.\cite{Convergence},
\begin{equation} \nonumber
    W_\rho (\boldsymbol{x},\boldsymbol{p}) = \int \frac{\dd^{2n} \boldsymbol{z}}{\pi^{2n}} \,  e^{(\boldsymbol{x}+i\boldsymbol{p}) \cdot \boldsymbol{z}^{\dagger}-(\boldsymbol{x}-i\boldsymbol{p})^{T} \!\! \cdot \boldsymbol{z}} \,  e^{-\frac{|\boldsymbol{z}|^{2}}{2}} \, \chi_\rho(\boldsymbol{z}).
\end{equation}
In the context of our work, two quantities of particular interest are the mean vector $\langle \hat{\boldsymbol{r}}\rangle_{\rho} = \tr(\hat{\boldsymbol{r}} \rho)$ and the covariance matrix $\boldsymbol{\gamma}$ with $\gamma_{ij} = \tr(\{ \Delta \hat{r}_i,\Delta \hat{r}_j \} \rho)/2$, where $\Delta \hat{r}_i = \hat{r}_i - \langle \hat{r}_i\rangle_{\rho}$ and $\{ \cdot, \cdot \}$ is the anticommutator.
Since $\rho$ is a quantum state, the $2n \times 2n$ real, symmetric covariance matrix satisfies the uncertainty principle $\boldsymbol{\gamma} + i \boldsymbol{\Omega} \geq 0$.

Bosonic Gaussian states will play a central role in the present work since they arise in the large-$n$ limit. They are those states whose characteristic function is given by a complex Gaussian, and are therefore entirely characterized by their mean vector and covariance matrix. In particular, a one-mode thermal state $\rho_{\text{th}}$ of average particle number $\tr(\hat{N} \, \rho_{\text{th}}) = N$ has zero mean vector and a covariance matrix $(N+\frac{1}{2})\mathds{1}_2$, where $\mathds{1}_k$ (for a positive integer $k$) denotes the identity matrix of dimension $k$. The characteristic function of the thermal state is thus written as the complex Gaussian
\begin{equation} \label{thermal}
\chi_{\rho_{\text{th}}}(\boldsymbol{z}) = e^{-N|\boldsymbol{z}|^{2}}.
\end{equation}
\subsection{Cushen and Hudson's quantum central limit theorem\label{sec:prelimQCLT}}
In the 70s, Cushen and Hudson introduced a quantum mechanical analogue of the well-known central limit theorem (CLT).
The latter states that, given $n$ independent and identically distributed random variables $X_{i}$ such that $\mathbb{E}[X_{i}]=0$, for $i=1,\ldots,n$, the random variable $Z_{n}$ defined as
\begin{equation} \label{eq:CLT}
    Z_{n} = \sum_{i=1}^{n} \frac{X_{i}}{\sqrt{n}}
\end{equation}
approaches a normally-distributed random variable with the same variance as the $X_{i}$ as $n$ goes to infinity.
Moving to bosonic systems, the summation of two random variables may be replaced by a ``quantum addition law'' acting on two quantum states $\rho_{1}$ and $\rho_{2}$ based on a beam splitter followed by partial tracing one of the two outputs. The beam-splitter unitary $\bs$ effects an energy-conserving linear coupling between modes $1$ and $2$ and is fully characterized by its effect on the mode operators, namely
\begin{equation} \label{eq:BogoliubovBS}
\begin{aligned}
\bs \, a_1 \, \bsd & = \sqrt{\eta} \, a_1 + \sqrt{1-\eta} \, a_2, \\
\bs \, a_2 \, \bsd & = - \sqrt{1-\eta} \, a_1 + \sqrt{\eta} \, a_2,
\end{aligned}
\end{equation}
where $\eta \in [0,1]$ is the transmittance of the beam splitter.
The quantum addition law is thus expressed as
\begin{equation} \label{eq:qadd2}
\sigma=\tr_{2}\left( \bs \rho_{1} \otimes \rho_{2} \bsd \right).
\end{equation}
Crucially, the analogy with the classical case (\textit{i.e.}, the summation of random variables) can be understood by realizing that the input mode operators  undergo a scaling followed by a summation, yielding the mode operator of the first output mode (in state $\sigma$) as expressed in the first equation of~\eqref{eq:BogoliubovBS}. Therefore, when $\eta=1/2$, the mode operators undergo the same operation as two random variables being transformed according to Eq.~\eqref{eq:CLT} for $n=2$. The analogy can be made even stronger in the following sense: consider the $\hat{x}$-marginals of the Wigner functions of the two input states, (with a slight abuse of notation)
\begin{equation} \nonumber
    W_{\rho_1}(x) = \int \dd p \, W_{\rho_1}(x,p), \quad W_{\rho_2}(x) = \int \dd p \, W_{\rho_2}(x,p).
\end{equation}
Here, $W_{\rho_1}(x)$ and $W_{\rho_2}(x)$ are proper (nonnegative) probability distributions whose associated random variables can be shown to undergo the exact transformation of Eq.~\eqref{eq:CLT} for $n=2$, yielding the $\hat{x}$-marginal of the Wigner function of the output state $\sigma$.

This quantum addition law can be easily generalized to $n$ modes by considering a so-called unbiased linear interferometer described by a unitary operator $\hat{U}$ acting on the tensor product of $n$ density matrices. Equivalently, this transformation can be characterized in phase space by a unitary matrix $U \in \mathbb{C}^{n\times n}$ acting as $\boldsymbol{a} \mapsto U \boldsymbol{a}$. The interferometer being unbiased, the first row elements of $U$ satisfy $U_{1,j} = 1/\sqrt{n}$ for all $j=1,\cdots,n$. One may then define an $n$-mode quantum addition law, acting on $n$ state $\rho_i$, for $i=1,\cdots,n$, as
\begin{equation} \label{eq:Qsum}
    \sigma_n = \tr_{\neg 1} \left( \hat{U} \bigotimes_{i=1}^{n} \rho_i \hat{U}^{\dagger} \right),
\end{equation}
where the partial trace is performed over all output modes but the first (see Fig.~\ref{Fig.1}). In that case, the output position and momentum operators are respectively given by the summations
\begin{equation}
    \tilde{x}_n = \sum_{i=1}^{n} \frac{x_i}{\sqrt{n}}, \quad \text{and} \quad \tilde{p}_n = \sum_{i=1}^{n} \frac{p_i}{\sqrt{n}},
\end{equation}
which are reminiscent of Eq.~\eqref{eq:CLT}.
The quantum central limit theorem (QCLT) derived by Cushen and Hudson predicts that the asymptotic form of $(\tilde{x}_n, \tilde{p}_n)$ of a single output mode obtained from the unbiased interference of $n$ identical states $\rho_i = \rho$ (for $i=1,\cdots,n$) according to the quantum addition law of Eq.~\eqref{eq:Qsum}, converges in distribution to a normal distribution~\cite{Central} (see Fig.~\ref{Fig.1}).
An equivalent result can be expressed as follows~\cite{Convergence}.

\begin{theo}\label{theo:qclt}
Given a state $\rho$ such that $\langle \hat{x}\rangle_{\rho}=\langle \hat{p}\rangle_{\rho}=0$, let $\rho_{\mathrm{G}}$ be the Gaussian state having the same covariance matrix as $\rho$. The output state $\sigma_{n}$ of the  $n$-mode unbiased interferometer with input $\rho^{\otimes n}$ satisfies
\begin{equation}
   \lim_{n\to \infty} \lVert\sigma_{n} - \rho_{\mathrm{G}}\rVert_{2}=0.
\end{equation}
\end{theo}
\noindent The statement can be proven by noticing that the pointwise convergence of the characteristic function can be translated into the theorem above using the quantum Plancherel relation~\cite{Convergence}:
\begin{equation}
    \lVert\sigma - \rho\rVert_{2}^{2}=\int \frac{\dd^{2n} \boldsymbol{z}}{\pi^{2n}}|\chi_{\sigma}(\boldsymbol{z})-\chi_{\rho}(\boldsymbol{z})|^{2}.
    \label{eq:plancherel}
\end{equation}
In particular, Theorem~\ref{theo:qclt} implies that if the initial covariance matrix of $\rho$ is proportional to the identity, the state $\sigma_n$ converges to a thermal Gaussian state. In the special case where  $\rho=|1\rangle\langle 1|$ is a single photon, $\sigma_{n}$ converges to a thermal state with an average particle number $N=1$, namely 
\begin{equation}
\sigma=\frac{e^{-\beta\hat{N}}}{\tr(e^{-\beta\hat{N}})},
\end{equation}
with $\beta=\ln(1+1/N)$.
Hence, the corresponding particle number distribution $p^{(\text{quant})}$, defined as the probability of seeing $m$ photons in the asymptotic state, i.e., $p_{m}^{(\text{quant})} = \lim_{n\to \infty}\langle m|\sigma_{n}|m \rangle$, follows a geometric distribution with a parameter $N/(1+N)$. This explains in a precise way the thermalization of a subsystem which is part of a bigger system evolving unitarily. 

This description of thermalization, however, is only valid when photons are fully indistinguishable. In realistic scenarios, the photonic states entering an interferometer often have differences in their internal degrees of freedom due, for example, to time delays or polarization differences. Even if the interferometer is oblivious to the internal degrees of freedom, such differences fundamentally affect the output photon statistics~\cite{Tichy2015}. Before we go into a detailed description of the generalized QCLT with partially distinguishable internal states $\rho_i$, let us consider the opposite extreme case where $\rho_i$ represent fully distinguishable single photons, i.e., $\tr(\rho_i \rho_j)= 0$, $\forall i\neq j$. The distinguishability of the photons fully suppresses quantum interference and the output statistics can be computed classically. In this setup, the probability that any of the input photons ends up in the first mode is $\frac{1}{n}$, hence the probability of measuring $m$ particles in the first mode is given by the  binomial distribution
\begin{equation}
p_{m}^{(\text{class})}=\binom{n}{m}(1/n)^{m}(1-1/n)^{n-m}\xrightarrow[n \to \infty]{}\frac{e^{-1}}{m!}.
\end{equation}
In the large-$n$ limit, it takes the form of a Poisson distribution with an average equal to 1, which is significantly different from the geometric distribution obtained for indistinguishable photons. These considerations motivate the development of a full description of the QCLT with partially distinguishable states, allowing us not only to predict  the probability distributions $p_m^{(\text{quant})}$ and  $p_{m}^{(\text{class})}$, but also the resulting photo-counting statistics for more complex partial distinguishability scenarios.   
 
\subsection{Partial distinguishability in bosonic systems}
To introduce the concept of partially distinguishable bosons, we begin with the simplest case, which describes two photons entering two different input arms of a beam splitter. The initial state can be expressed as~\cite{GSO}:
\begin{equation}\label{2-photon-input}
|\Psi_{\text{in}}\rangle=a_{1,\phi_{1}}^{\dagger}a_{2,\phi_{2}}^{\dagger}|vac\rangle.
\end{equation}
Here, the operator $a_{i,\phi_{j}}^{\dagger}$ generates a photon in spatial mode $i$ with internal degree of freedom (such as frequency or polarization) described by a wavefunction $|\phi_{j}\rangle$, and $\ket{vac}$ represents the vaccuum state. While the states $\ket{\phi_1}$ and  $\ket{\phi_2}$ are in general non-orthogonal, we can express them in a two-dimensional orthonormal basis $\{\ket{u}\}$ as 
\begin{align}
|\phi_{1}\rangle=\sum_{v=1}^{2}c_{1,u}|u\rangle, ~~~~~ 
 \ |\phi_{2}\rangle=\sum_{v=1}^{2}c_{2,u}|u\rangle .
\end{align}
Consequently, we can express the initial state as:
\begin{equation}
|\Psi_{\text{in}}\rangle={\bigg (}\sum_{u,v}c_{1,u}\, c_{2,v}\, a^{\dagger}_{1,u}\, a^{\dagger}_{2,v}{\bigg )}|vac\rangle, 
\end{equation}
where the operators $a_{i,u}$ obey the usual commutations relations $[a_{i,u}, a^{\dagger}_{j,v}]= \delta_{i,j}\delta_{u,v}$. It is well known that if this state goes through a 50-50 beam-splitter which couples only the spatial modes, the probability of observing one photon in each output mode (\textit{i.e.}, the coincidence probability) depends on the overlap $|\braket{\phi_1|\phi_2}|^2$. 

More generally, the linear interference of multiple partially distinguishable photons is usually described by a unitary which acts on creation operators as 
\begin{equation} \label{eq:qaddn}
    \hat{U} a^{\dagger}_{i, u} \hat{U}^\dagger= \sum_{j} U_{ij} a^{\dagger}_{j, u},  
\end{equation}
coupling the different spatial modes $i$ and $j$, but leaving internal degrees of freedom $u$ unchanged.  Assuming detectors can only count the total number of photons in a given spatial mode (independently of their internal degrees of freedom), the photo-counting statistics of such an experiment depends only on the unitary $U_{ij}$ and the distinguishability matrix $S$, defned as 
\begin{equation}
S_{i,j}=\langle \phi_{i}|\phi_{j}\rangle, 
\end{equation}
which is the Gram matrix associated with the pairwise overlaps of the internal states $\ket{\phi_j}$~\cite{Tichy2015}. While the specific form of the state  describing the internal degrees of freedom is not crucial for our discussion, it is important to note that with $n$ particles, we can always choose an orthonormal basis of dimension at least $d\leq n$  to describe all internal states $\ket{\phi_i}$, where $d=\text{rank}(S)$ is the dimension of the Hilbert space spanned by the states $\{\ket{\phi_i}\}$.

\section{A quantum central limit theorem with partial distinguishability}\label{sec:A quantum central limit theorem with partial distinguishability}

\subsection{Unbiased interference of partially distinguishable bosons}

\begin{figure}[t!]
    \centering
    \includegraphics[scale=0.34]{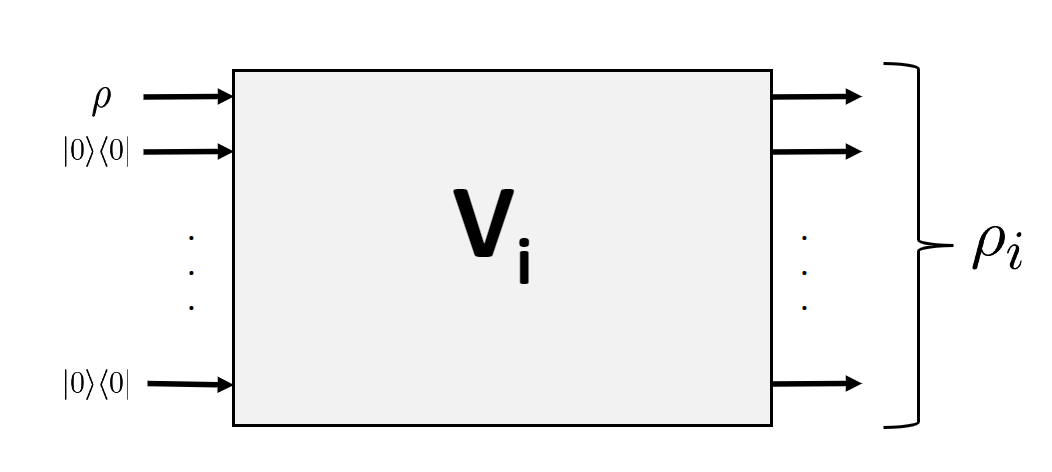}
    \caption{Process used to model the partial distinguishability of the input states $\ket{\phi_i}$ as described by Eq. (\ref{state-gen}). This can be viewed as a unitary $V_i$ acting on the internal modes associated with each spatial mode, starting from an input state~$\rho$ occupying a first (reference) internal mode, while the other internal modes are initially in the vacuum state. The resulting state $\rho_i$ is defined over the $d$ internal modes associated with each spatial mode. The product state $\prod_{i=1}^n \rho_i$ then serves as an input of the unbiased interferometer.  }
    \label{fig:V_i}
\end{figure}

We start with the same setting as in the QCLT, namely, with $n$ identical sates $\rho$ satisfying $\braket{\hat{x}}_{\rho}=\braket{\hat{p}}_{\rho}=0$, one in each of the $n$ spatial modes. We assume that the internal degrees of freedom of the photons in  state $\rho$ are described by a reference wavefunction, which we denote as $\ket{1}$. In other words, the state $\rho$ in spatial mode $i$ can be written solely as functions of the creation/annihilation operators $a^{\dagger}_{i,1}$ and $a_{i,1}$ acting on vacuum. Partial distinguishability is introduced by the action of a set of unitaries $V_i$ acting solely on the internal degrees of freedom via the linear transformation 
\begin{equation}\label{p.d. unitary}
\hat{V}_{i}a_{i,1}^{\dagger}\hat{V}_{i}^{\dagger}=\sum_{u=1}^{d}(V_{i})_{1,u}a_{i,u}^{\dagger}=\sum_{u=1}^{d}c_{i,u}a_{i,u}^{\dagger}, 
\end{equation}
for some choice of an orthonormal basis $\{|u\rangle\}$ describing the internal degrees of freedom. Throughout this work, we choose the minimum possible dimension for the basis describing these internal states, \textit{i.e.}, $d=\text{rank}(S)$. The state in spatial mode $i$ is thus now described by a $d$-mode density matrix 
\begin{equation}\label{state-gen}
\rho_{i}= \hat{V}_{i} \, \rho \,  \hat{V}_{i}^{\dagger} ,
\end{equation}
as depicted in Fig.~\ref{fig:V_i}.
We note that the coefficients $c_{i,u}$ can be related to the Gram matrix, since
\begin{equation}
\sum_{u=1}^{d}c_{i,u}^{*}c_{j,u}=\langle \phi_{i}|\phi_{j}\rangle=S_{i,j} \implies S=C C^{\dagger} ,
\end{equation}
where we have defined the $n\times d$ matrix $C$, whose entries are $c_{i,u}$.
This model for the input state encompasses as particular cases several models of partial distinguishability considered in the literature, such as single photons in different pure internal states (when $\rho$ is a single photon state) \cite{Tichy2015} or  particular models of partial distinguishability for Gaussian Boson sampling~\cite{shi2022part_dist_GBS} (when $\rho$ is a squeezed state). 

The unbiased interference of the states $\rho_i$ is now described by an interferometer $\hat{U}$ acting only on the spatial degrees of freedom, so that the $d$-mode output state occupying the first spatial mode is described by mode operators given by the unbiased sums 
\begin{equation}\label{unitary transform}
    b^{\dagger}_{1, u} = \sum_{j=1}^{n} \frac{a^{\dagger}_{j, u}}{\sqrt{n}}.
\end{equation}
Similarly to the standard QCLT considered by Cushen and Hudson, which can be recovered in the case where all unitaries $V_i$ act as the identity, we are interested in the description of the asymptotic state $\sigma_n$, resulting from the unbiased quantum addition of states with different internal degrees of freedom (see Eq.~\ref{eq:Qsum}). Here, $\sigma_n$ is the reduced density matrix of the first spatial mode at the output of the interferometer (represented in Fig.~\ref{Fig.1}), hence it is a $d$-mode bosonic quantum state defined over the Hilbert space encompassing all internal degrees of freedom.
\subsection{Asymptotic state}
In order to state our main result in a concise way, it is useful to define the ($d$-mode) state of the uniform mixture of all input internal states $\rho_i$, that is,
\begin{equation}
    \bar{\rho}= \frac{1}{n}\sum_{i=1}^n  \rho_i.
    \label{eq:rho_average}
\end{equation}
In this equation, it is understood that we are mixing the states $\rho_i$ describing the internal degrees of freedom, as if they were all occupying the same spatial mode.
\begin{theo}\label{theo:main}
Let $\bar{\rho}_{\mathrm{G}}$ be the multimode Gaussian state having the same covariance matrix as $\bar{\rho}$. The multimode output state $\sigma_{n}$ of the unbiased interferometer with partially distinguishable inputs satisfies
\begin{equation}
   \lim_{n\to \infty} \lVert\sigma_{n} - \bar{\rho}_{\mathrm{G}}\rVert_{2}=0.
\end{equation}
\end{theo}
\noindent A detailed derivation of the result is given in Appendix~\ref{Proof of the main}. As in the original proof of the bosonic QCLT, the main ingredient is to derive the characteristic function of the asymptotic state $\sigma_n$, that is,
\begin{equation} \label{eq:chi1}
    \chi_{\sigma_{n}}(\boldsymbol{z}_1)= \tr\left(\sigma_n \, e^{\sum_{u}z_{1,u}b_{1,u}^{\dagger}}e^{\sum_{u}-z_{1,u}^{*}b_{1,u}}\right), 
\end{equation}
where we define the $d$-dimensional vector of complex variables $\boldsymbol{z}_i=(z_{i,1}, ..., z_{i,d})$. By Taylor expanding this function to second order, we show it is approximated by a multivariate Gaussian since it satisfies 
\begin{multline}
\ln \chi_{\sigma_{n}}(\boldsymbol{z}_1) = -\boldsymbol{z}_1^{\dagger}\frac{C^{\dagger}C}{n}\boldsymbol{z}_1\langle a^{\dagger}a\rangle  - \frac{1}{2}\boldsymbol{z}_1^{T}\frac{C^{T}C}{n}\boldsymbol{z}_1\langle a^{\dagger}a^{\dagger}\rangle\\ -\frac{1}{2}\boldsymbol{z}_1^{\dagger}\frac{(C^{T}C)^*}{n}\boldsymbol{z}_1^{*}\langle aa\rangle+O(n^{-\frac{1}{2}}),\label{eq:main_result}
\end{multline}
for large $n$. Hence, the asymptotic state $\sigma_n$ tends to a multimode Gaussian state. In the above, the expectation values are evaluated with respect to the state $\rho$, i.e., $\langle \hat{A}\rangle=\tr(\hat{A}\rho)$, and can be expressed as a function of the covariance matrix $\tau$ of the single-mode input state $\rho$ via
\begin{equation}
    \begin{aligned}
        \braket{ a a } & = \frac{ \tau_{11}- \tau_{22}+ i \tau_{12}}{2}, \\
   \braket{ a^\dagger a } & = \frac{\tau_{11}+ \tau_{22}}{2}.
    \end{aligned}
\end{equation}
To better understand this result, a phase-space representation in terms of position and momentum variables can be provided, which follows from the change of variables
\begin{equation} \label{eq:ztoxp}
\Re{(z_{1, u})}=x_u \ \ , \ \ \Im{(z_{1,u})}=p_u \implies \boldsymbol{z_1},\boldsymbol{z_1}^{*}\to \boldsymbol{x},\boldsymbol{p}.
\end{equation}
By combining the phase-space variables in a vector $\boldsymbol{y}=(x_{1},...,x_{d},p_{1},...,p_{d})$, the final characteristic function can be expressed as:
\begin{equation} \label{eq:eq:chi2}
\chi_{\sigma_{n}}(\boldsymbol{y})=e^{-\boldsymbol{y}^{T}\gamma\boldsymbol{y}}+O(n^{-\frac{1}{2}}).
\end{equation}
In what follows, with a slight abuse of notation, when $\chi_{\sigma_{n}}$ has argument $\boldsymbol{z}_1$, it will refer to Eq.~\eqref{eq:chi1}, while when it has argument $\boldsymbol{y}$, it will refer to Eq.~\eqref{eq:eq:chi2}.
In the above, $\gamma$ depends on the elements of the Gram matrix $S$ (through elements $c_{i,u}$) and the second moments of the initial states $\rho$. Specifically, $\gamma$ can be written as:
\begin{equation} \label{eq:covariance_sigma}
    \begin{aligned}
        \gamma=\mathds{1}_2\otimes \frac{C^{\dagger}C}{n}\langle{a^{\dagger}a}\rangle & + \sigma_{z}\otimes \Re{\bigg [}\frac{C^{T}C}{n}\langle a^{\dagger}a^{\dagger}\rangle{\bigg ]} \\
        & +iJ \otimes\Im{\bigg [}\frac{C^{T}C}{n}\langle a^{\dagger}a^{\dagger}\rangle{\bigg ]},
    \end{aligned}
\end{equation}
where  $J$ is the anti-diagonal identity matrix of dimension~$2$, while $\sigma_{z}$ is the $z$-component Pauli matrix.
\subsection{Physical interpretation}
To interpret the previous result it is useful to diagonalize the quadratic form appearing in Eq.~\eqref{eq:main_result}. We note, however, that $C$ is in general a complex matrix and thus $C^{T}C$ and $C^{\dagger}C$ are not diagonal in the same basis. For this reason, we focus on two situations of physical interest which allow us to give a simple physical interpretation of the form of the asymptotic multimode Gaussian state, which explicitly depends on the non-zero eigenvalues of the $S$ matrix. For simplicity, we will denote the asymptotic state as $\sigma= \lim_{n\to\infty} \sigma_{n}$.
\subsubsection{Input Fock states and thermal states}\label{sec:Fock_thermal}
We focus first on the special case where the input states $\rho$, before undergoing the transformations $V_i$ accounting for partial distinguishability, are such that $\braket{x^2}= \braket{p^2}$ and $\braket{\{x, p\}}=0$. In other words, we assume that $\rho$ admits a covariance matrix $\tau = r \mathds{1}_2$, describing situations with Fock states $\ket{r}$ or thermal states at the input. In this case, Eq.~\eqref{eq:main_result} can be simplified,  taking the form 
\begin{equation}
 \chi_{\sigma}(\boldsymbol{z}_1)=\exp\left(-\boldsymbol{z}_1^{\dagger}\Gamma\boldsymbol{z}_1\langle a^{\dagger}a\rangle\right).\label{eq:main_result_diag}
\end{equation}
Here, we have introduced the $d\times d$ matrix $\Gamma$ whose elements are defined as
\begin{equation}\label{density}
\Gamma_{u v }=\frac{1}{n}\sum_{i=1}^n\braket{u|\phi_{i}}\braket{\phi_{i}|v}=\frac{(C^{\dagger}C)_{uv}}{n},  
\end{equation}
hence, $ \Gamma =  C^{\dagger}C /n$. Remarkably, this means that the matrix $\Gamma$ has the same $d$ non-zero eigenvalues as the (rescaled) distinguishability matrix $S/n$~\cite{Jozsa_2000}, which we denote by  $\{\lambda_{u}\}_{u=1,\ldots,d}$.
By choosing a basis $\{\ket{u'}\}$ for the internal degrees of freedom that diagonalizes $\Gamma$, we can rewrite Eq.~\eqref{eq:main_result_diag} as 
\begin{equation}
\chi_{\sigma}(\boldsymbol{z}'_1)=\prod_{u=1}^{d}e^{-\lambda_{u} \, r \, |z'_{1,u}|^{2}}, 
\end{equation}
which is the characteristic function of a generalized Gibbs state, that is, a product of thermal states (one for each dimension of the internal Hilbert space). It is interesting to observe that a natural orthogonal basis for the internal degrees of freedom appears in which to describe the final state via a generalized Gibbs state. The eigenbasis of $\Gamma$ is, in fact, the eigenbasis of the single-particle quantum state described by the uniform mixture $\bar{\rho}$ of states $\ket{\phi_i}$ (see Eq.~\eqref{density}). The eigenvalues of state $\bar{\rho}$ -- and thus of the (rescaled) distinguishability matrix $S/n$ --
acquire an important physical role as they  define the different ``temperatures" associated with the different orthogonal degrees of freedom. As a consistency check, we note that when all states $\ket{\phi_i}$ are identical, we have that $\Gamma$ has a single non-zero eigenvalue $\lambda=1$, thus we recover the prediction from the QCLT for indistinguishable bosons.
\subsubsection{Real distinguishability matrices}
While the imaginary part of the distinguishability matrix $S$ leads to interesting photonic interference phenomena~\cite{shchesnovich2018collective, Menssen2017} and has measurable consequences in many different fields of quantum information~\cite{qloc2024unitaryinvariant}, the most common source of partial distinguishability in photonic experiments comes from time-delays which can be modelled by real distinguishability matrices~\cite{Menssen2017}. In particular, this means that we can choose a basis in which to represent the internal states $\ket{\phi_i}$ which is such $C$ has only real entries, implying that $\Gamma= C^T C /n$. In this scenario, we can simplify Eq.~\eqref{eq:covariance_sigma}, which becomes
\begin{equation}
    \gamma = \tau\otimes \Gamma,
\end{equation}
a tensor product of the two-dimensional covariance matrix of $\rho$ with the matrix $\Gamma$ which contains the information about partial distinguishability. Similarly to Sec.~\ref{sec:Fock_thermal}, the diagonalization of $\Gamma$ allows us to write the characteristic function as 
\begin{equation}
\chi_{\sigma}(\boldsymbol{z}'_1)=\prod_{u=1}^{d}e^{-\lambda_{u}(|z'_{1,u}|^{2}\braket{a^\dagger a}- \frac{1}{2}(z_{1, u})^2\braket{a^\dagger a^\dagger }- \frac{1}{2}(z^*_{1, u})^2\braket{a a})}, 
\end{equation}
corresponding to a product of squeezed states over the different internal degrees of freedom. 
\section{Particle number distribution}\label{sec:GF}
We now turn our focus to the particle-number distribution associated with the asymptotic output state $\sigma$, which we denote by $\{ p_m \}_{m \in \mathbb{N}}$. In other words, $p_m$ is the total probability of finding $m$ particles in the first output spatial mode of the interferometer when counting all internal degrees of freedom (and in the limit of $n$ going to infinity).
Indeed, it is important to note that the individual internal degrees of freedom of a system are not easily accessible experimentally (for example, the time resolution may be limited). Instead, one may have access to observables that are independent of them, such as particle counting in a given spatial mode.

To characterize the distribution $\{ p_m \}_{m \in \mathbb{N}}$, it is useful to introduce what we will refer to as the particle-number generating function, which can be defined as follows~\cite{Counting}:
\begin{equation}\label{generating-function}
G(\beta)=\int \frac{\dd^{2d} \boldsymbol{y}}{\pi^{d}(1-\beta)^{d}}\, e^{-\frac{|\boldsymbol{y}|^{2}}{1-\beta}}\, \chi_{\sigma}(\boldsymbol{y}),
\end{equation}
with $\beta \in \mathbb{R}$.
The generating function $G(\beta)$ encompasses all the information on the particle number distribution. Indeed, the probability of measuring $m$ photons in the first output mode of the interferometer (regardless of the internal degree of freedom) can then be recovered as
\begin{equation}
p_{m}=\frac{1}{m!}\frac{\partial^{m}}{\partial \beta^{m}}G(\beta){\bigg |}_{\beta=0}.
\end{equation}
In the limit $n\to \infty$, the Gaussian form of the characteristic function of the asymptotic state $\sigma$ leads to the following simple expression for the generating function
\begin{equation}
G(\beta)=\frac{1}{\sqrt{\det \left(\mathds{1}_{2d}+(1-\beta)\gamma\right)}}.
\end{equation}
If one chooses the input state $\rho$ to have a simple covariance matrix propositional to the identity matrix ($\tau=r \, \mathds{1}_2$), the above relation can be further simplified to
\begin{equation} \label{eq:genId}
G(\beta)=\frac{1}{\det(\mathds{1}_d+r(1-\beta)\Gamma)}.
\end{equation}
This encompasses the cases of thermal states and Fock states, the latter being of particular interest. Notably, the extreme cases discussed in Section~\ref{sec:prelimQCLT} for the single-photon inputs can easily be recovered from Eq.~\eqref{eq:genId}:
\begin{itemize}
\item The quantum case (fully indistinguishable photons) is obtained when $\text{rank}(S)=1$, implying a single non-zero eigenvalue $\lambda_u=1$. This results in
\begin{equation}
p_{m}^{(\text{quant})}=\frac{1}{m!}\frac{\partial^{m}}{\partial \beta^{m}}\frac{1}{2-\beta}{\bigg |}_{\beta=0}=\frac{1}{2^{m+1}},
\end{equation}
which is indeed a geometric distribution with mean value $1$, as expected.
\item The classical case (fully distinguishable photons) is obtained for $S=\mathds{1}_n$, in which case the eigenvalues of $\Gamma$ are $\{ \frac{1}{n},...,\frac{1}{n}\}$. This case requires careful treatment due to each eigenvalue tending to zero while their number diverges. In the large-$n$ limit, we obtain as generating function
\begin{equation}
G_{|1\rangle\langle1|}^{(\text{class})}(\beta)={\bigg (}\frac{1}{1+\frac{1-\beta}{n}}{\bigg )}^{n}\sim e^{\beta-1}.
\end{equation}
Consequently, we have
\begin{equation}
p_{m}^{(\text{class})}=\frac{1}{m!}\frac{\partial^{m}}{\partial \beta^{m}}e^{\beta-1}{\bigg |}_{\beta=0}=\frac{e^{-1}}{m!},
\end{equation}
which is indeed a Poisson distribution with mean value $1$, as expected.
\end{itemize}
In the more general case where the covariance matrix of the initial state $\rho$ is of the form $r \, \mathds{1}_2$, the particle-number distribution can be shown to satisfy the following relation
\begin{equation} \label{eq:recur}
 p_{m}=\frac{1}{m}\sum_{l=0}^{m-1}{\bigg (}p_{l}\sum_{u=1}^{d}{\bigg [}\frac{r\lambda_{u}}{1+r\lambda_{u}}{\bigg ]}^{m-l}{\bigg )}, \quad \forall \, m \geq 1.
\end{equation}
We point the interested reader to Appendix~\ref{Recursive} for a proof of the above.
It is interesting to note that computing the probabilities $p_{m}$ can be done in polynomial time by means of this recursion relation only for a Gram matrix satisfying $\text{rank}(S)=d<\infty$. The infinite dimensional case can be treated by carefully taking the limit $n\to \infty$ into account.

For instance, let us focus on a simple model, characterized by a Gram matrix whose elements are defined as 
\begin{equation}
    \tilde{S}_{i,j}(x)=\delta_{i,j}+x(1-\delta_{i,j}), \quad \forall i,j=1,2,\ldots,
\end{equation}
for some $x \in [0,1]$. This interpolation model of partial distinguishability was, for instance, considered in Refs.~\cite{Distinguishability, renema_dist}.
The rank of $\tilde{S}(x)$ is finite only for fully indistinguishable bosons (for $x=1$), but is infinite otherwise. For finite $n$, the matrix $\tilde{S}(x)/n$ has maximum and minimum eigenvalues respectively given by:
\begin{equation}
    \lambda_\text{max}=\frac{1+(n-1)x}{n} \quad \text{and} \quad \lambda_\text{min}=\frac{1-x}{n},
\end{equation}
with $\lambda_\text{min}$ having multiplicity $n-1$. Let us consider the simplest example in which we have one single particle per input. Computing $p_{0}(x)$ can be done by taking the limit
\begin{equation}
    p_{0}(x)=\lim_{n\to\infty}\prod_{u=1}^{n}\frac{1}{1+\lambda_{u}}=\frac{e^{x-1}}{1+x}.
\end{equation}
Using the recursion relation of Eq.~\eqref{eq:recur} and carefully taking the limit, we can provide a closed form for the probability distribution as:
\begin{equation}
    p_{m}(x)=\frac{e^{x-1}}{1+x}(1-x)^{m}\sum_{i=0}^{m}\frac{1}{(m-i)!}{\bigg (}\frac{x}{1-x^{2}}{\bigg )}^{i}.
\end{equation}
This is the convolution between a Poisson distribution with mean value $1-x$ and a geometric distribution with mean value $x$, meaning that the convolution itself has a mean value $1$, as expected.
This result has a simple interpretation as follows. On average, a fraction $x$ of the total number of particles occupies the same internal state and so they interfere with each other leading to a geometric distribution. In contrast, a fraction $1-x$ are fully distinguishable and their behavior follows purely classical statistics, leading to the Poisson distribution. The total number of particles will be the sum of these two contributions and so its corresponding probability distribution will be the convolution of the geometric and Poisson distributions mentioned above.

Knowing the generating function also allows us to compute the moment generating function:
\begin{equation}
M(\beta)=G(e^{\beta}) \implies \langle \hat{N}^{k}\rangle_{\boldsymbol{p}}=\frac{\partial^{k}}{\partial \beta^{k}}M(\beta){\bigg |}_{\beta=0}
\end{equation}
Assuming again $\tau=r \, \mathds{1}_2$, it can be proven that the mean photon number as its variance are given by
\begin{equation}
\langle \hat{N}\rangle=r \ \ , \ \ \text{Var}(\hat{N})=r(1+\tr(\Gamma^{2})),
\label{eq:av_and_var}
\end{equation}
respectively. In particular, it is interesting to remark that the variance of the photon number distribution increases with indistinguishability. In fact, the quantity $\tr(\Gamma^{2})$ is the purity of the uniform mixture $\bar{\rho}$ of internal states $\ket{\phi_i}$, which can be seen as a measure of indistinguishability since 
\begin{equation}
    \tr(\Gamma^{2})= \frac{1}{n^2}\sum_{i,j} |\braket{\phi_i| \phi_j}|^2, 
\end{equation}
gets closer to 1 by increasing the overlap between the different wavefunctions representing the internal degrees of freedom. This quantity is minimized for fully distinguishable photons, \textit{i.e.}, orthogonal states $\ket{\phi_i}$, and maximized when all states are the same. Hence, observing a photon number variance of the asymptotic state $\sigma_n$ that is close to the maximum value $2 r $ can be seen as a signature of high indistinguishability of the input states. 
\section{Conclusion}\label{sec:conclusions}
Our work generalizes the Cushen-Hudson QCLT by explicitly taking into account the fact that bosons may have differences in their internal degrees of freedom, which are left invariant by the interferometer but play a central role in thermalization. Using characteristic function techniques, we showed that the unbiased interference of a large number of partially distinguishable bosons converges to a multimode Gaussian state
defined over the Hilbert space describing these internal degrees of freedom. For experiments with input Fock states, we observe thermalization towards a generalized Gibbs state. The number of temperatures defining this state is given by the dimension of the Hilbert space spanned by the different internal states, while the different values of the temperatures are given by the non-zero eigenvalues of the distinguishability matrix $S$. While the use of $S$ has been ubiquitous in the modelling of linear interference between partially distinguishable bosons, it was previously unnoticed that its eigenvalues play an important physical role. 

In addition, even though internal degrees of freedom may not be directly accessible, our results suggest that the photon number distributions (ignoring the internal degrees of freedom) of reduced subsystems of the output state of an interferometer contain valuable information of about bosonic indistinguishability. This information may be used to diagnose experimental imperfections in bosonic experiments.  Moreover, we believe that our work may be generalized to other interesting settings, such as Haar random interferometers \cite{bosonsampling} or non-interacting lattice Hamiltonians \cite{cramer2008lattice}, and lead to a better understanding of the role of partial distinguishability in equilibration phenomena which could be observed via current photonic and atomic boson sampling experiments.  
\begin{acknowledgments}
We thank Antoine Restivo for many valuable discussions. M.R. and N.J.C. acknowledge  funding from the European Union’s Horizon 2020 research and innovation programme under Marie
Skłodowska-Curie grant agreement No. 956071 (AppQInfo). M.G.J. acknowledges support from the Fonds de la Recherche Scientifique – FNRS (Belgium). During part of this project, L.N. was a senior postdoctoral fellow of the Fonds de la Recherche Scientifique – FNRS (Belgium). L.N. acknowledges support from FCT-Fundação para a Ciência e a Tecnologia (Portugal) via the Project No. CEECINST/00062/2018 and from the European Union's Horizon Europe research and innovation program under EPIQUE Project GA No. 101135288. N.J.C. acknowledges support from both the European Union and Fonds de la Recherche Scientifique – FNRS (Belgium) under project ShoQC within the ERA-NET Cofund in Quantum Technologies \mbox{(QuantERA)} program.
\end{acknowledgments}

\bibliography{apssamp}

\appendix

\section{Some proof and calculation details}
\subsection{Proof of the main theorem\label{Proof of the main}}
To prove Theorem~\ref{theo:main}, we start by computing the expression for the characteristic function of the state $\sigma_n$. Recall that the characteristic function of the state at the input of the unbiased interferometer can be written as
\begin{equation}
    \chi_{\otimes_{i=1}^{n}\rho_{i}}(\boldsymbol{z})=\prod_{i=1}^{n}\chi_{V_{i}\rho V_{i}^{\dagger}}(\boldsymbol{z}_{i}).
\end{equation}
Making use of Eqs.~\eqref{p.d. unitary} and~\eqref{unitary transform} to apply the unbiased interferometer, after tracing out over all the spacial modes except the first one, we obtain
\begin{equation}
\chi_{\sigma_{n}}(\boldsymbol{z}_{1})=\prod_{i=1}^{n} \chi_{\rho}{\bigg (}\frac{c_{i,1}z_{1,1}}{\sqrt{n}},...,\frac{c_{i,k}z_{1,k}}{\sqrt{n}}{\bigg )}.
\end{equation}
We can expand in series the single characteristic functions $\chi_{\rho}(\boldsymbol{z}_{1})$ appearing in the product as
\begin{equation}\label{single-m}
    \begin{aligned}
        \chi_{\rho}(\boldsymbol{z}_{1}) = 1 & - \frac{1}{2n}\sum_{u,v}2\langle b^{\dagger}_{i,u}b_{i,v}\rangle z_{1,u}z_{1,v}^{*} \\
        & + \langle b_{i,u}b_{i,v}\rangle z_{1,u}^{*}z_{1,v}^{*}+\langle b^{\dagger}_{i,u}b_{i,v}^{\dagger}\rangle z_{1,u}z_{1,v} \\
        & + O(n^{-\frac{3}{2}}).
    \end{aligned}
\end{equation}
The value of $\langle b^{\dagger}_{i,u}b_{i,v}\rangle $ can be computed as a function of the coefficients $c_{i,u}$ and $\langle a^{\dagger}a\rangle_{\rho}=\tr(a^{\dagger}a\rho) $. In particular, given Eq.~\eqref{p.d. unitary}, we have
\begin{equation}
    \begin{aligned}
        \langle b_{i,u}^{\dagger}b_{i,v}\rangle & = \tr{\bigg (}\sum_{u'}\sum_{v'}(V_{i})_{u,u'}(V_{i})_{v,v'}^{*}a_{i,u'}^{\dagger}a_{i,v'}\rho{\bigg )} \\
        & = \sum_{u'}\sum_{v'}(V_{i})_{u,u'}(V_{i})_{v,v'}^{*}\tr{\bigg (}a_{i,u'}^{\dagger}a_{i,v'}\rho{\bigg )} \\
        & = (V_{i})_{u,1}(V_{i})_{v,1}^{*}\tr{\bigg (}a_{i,1}^{\dagger}a_{i,1}\rho{\bigg )} \\
        & = c_{i,u}c_{i,v}^{*}\tr{\bigg (}a_{i,1}^{\dagger}a_{i,1}\rho{\bigg )}
    \end{aligned}
\end{equation}
where we have used the fact that
\begin{equation}
    \tr(a_{i,u'}^{\dagger}a_{i,v'}\rho)=\tr(a_{i,1}^{\dagger}a_{i,1}\rho)\delta_{u',1}\delta_{v',1}.
\end{equation}
The quantities $\langle b_{i,u}b_{i,v}\rangle,\langle b_{i,u}^{\dagger}b_{i,v}^{\dagger}\rangle$ can be computed in a similar way. At this point by substituting this in the previous equation we get:
\begin{multline}
    \chi_{\rho}(\boldsymbol{z}_{i})=1-\frac{1}{2n}\sum_{u,v}2\langle a^{\dagger}a\rangle c_{i,u}c_{i,v}^{*} z_{1,u}z_{1,v}^{*}+\\
    \langle aa\rangle 
    c_{i,u}^{*}c_{i,v}^{*}z_{1,u}^{*}z_{1,v}^{*}+\langle a^{\dagger}a^{\dagger}\rangle 
    c_{i,u}c_{i,v}
    z_{1,u}z_{1,v} + O(n^{-\frac{3}{2}})
\end{multline}
where we shorten the notation using $\tr(AB\rho)=\langle AB\rangle$. By combining this with eq.(\ref{single-m}) we get:
\begin{multline}
    \chi_{\sigma_{n}}(\boldsymbol{z}_{1})=1-\frac{1}{2n}\sum_{i}\sum_{u,v}2\langle a^{\dagger}a\rangle c_{i,u}c_{i,v}^{*} z_{1,u}z_{1,v}^{*}+\\
    \langle aa\rangle 
    c_{i,u}^{*}c_{i,v}^{*}z_{1,u}^{*}z_{1,v}^{*}+\langle a^{\dagger}a^{\dagger}\rangle 
    c_{i,u}c_{i,v}
    z_{1,u}z_{1,v} + O(n^{-\frac{1}{2}})
\end{multline}
In the limit $n \to \infty$ we can write:
    \begin{multline}
\lim_{n\to \infty}\ln \chi_{\sigma_{n}}(\boldsymbol{z}_1) = -\boldsymbol{z}_1^{\dagger}\frac{C^{\dagger}C}{n}\boldsymbol{z}_1\langle a^{\dagger}a\rangle+\\  - \frac{1}{2}\boldsymbol{z}_1^{T}\frac{C^{T}C}{n}\boldsymbol{z}_1\langle a^{\dagger}a^{\dagger}\rangle -\frac{1}{2}\boldsymbol{z}_1^{\dagger}\frac{(C^{T}C)^*}{n}\boldsymbol{z}_1^{*}\langle aa\rangle.
\end{multline}
It can be seen that the covariance matrix defining this multimode Gaussian state is the same as that of the average state $\bar{\rho}$ defined in Eq.~\eqref{eq:rho_average}. This pointwise convergence of the characteristic function can be translated into a convergence in two-norm to the asymptotic state, via the quantum Plancherel relation from Eq.~\eqref{eq:plancherel}. This completes the proof of Theorem~\ref{theo:main}. 
\subsection{Recursive relation for the particle-number distribution of the asymptotic state\label{Recursive}}
To prove the structure of the recursive form of the probability distribution, let us start by recalling the Jacobi's formula, for an invertible matrix $A(t)$:
\begin{equation}
    \frac{\dd}{\dd t}\det(A(t))=\det(A(t))Tr{\bigg (}A(t)^{-1} \frac{\dd}{\dd t}A(t) {\bigg )}
\end{equation}
And thus we have that:
\begin{equation}
    \frac{\dd}{\dd t}\frac{1}{\det(A(t))}=\frac{Tr{\bigg (}A(t)^{-1} \frac{\dd}{\dd t}A(t) {\bigg )}}{\det(A(t))}
\end{equation}
We also recall:
\begin{equation}\label{derivatives of inverse}
    \frac{\dd}{\dd t}Tr(A(t))=-Tr{\bigg (}A(t)^{-1} \frac{\dd}{\dd t}A(t)  A(t)^{-1}{\bigg )}
\end{equation}
By definition, for an input state $\rho$ with diagonal covariance matrix, such as a Fock state $\rho=|r\rangle\langle r|$, we have:
\begin{equation}
    p_{0}=G(\beta){\bigg |}_{\beta=0}=\frac{1}{\det(\mathds{1}_d+r\Gamma)}
\end{equation}
By following the definition and using Jacobi's formula we have that:
\begin{equation}
    \frac{\partial}{\partial\beta}G(\beta)=G(\beta)Tr(r\Gamma(\mathds{1}_d+(1-\beta)r\Gamma)^{-1})
\end{equation}
We recall that $\Gamma$ is a $d\times d$ full rank matrix, whose eigenvalues are the non-zero eigenvalues of the distinguishability matrix $S$, and so $\Gamma$ is invertible. We will define $f(\beta)=Tr(r\Gamma(\mathds{1}_d+(1-\beta)r\Gamma)^{-1})$ and thus we can use the product rule for derivatives:
\begin{equation}
    \frac{\partial^{m}}{\partial\beta^{m}}{\bigg [}G(\beta)f(\beta){\bigg ]}=\sum_{k=0}^{m}\binom{m}{k}\frac{\partial^{m-k}}{\partial\beta^{m-k}}G(\beta)\frac{\partial^{k}}{\partial\beta^{k}}f(\beta).
\end{equation}
We further notice that:
\begin{equation}
    p_{m+1}=\frac{1}{(m+1)!}\frac{\partial^{m}}{\partial\beta^{m}}{\bigg [}G(\beta)f(\beta){\bigg ]}_{\beta=0}.
\end{equation}
We can combine the two equation above to obtain:
\begin{equation}
    p_{m+1}=\frac{1}{m+1}\sum_{k=0}^{m}p_{m-k}\frac{1}{k!}\frac{\partial^{k}}{\partial\beta^{k}}f(\beta){\bigg |}_{\beta=0}.
\end{equation}
The derivatives of $f(\beta)$ can be computed using eq.(\ref{derivatives of inverse}):
\begin{equation}
    \frac{1}{k!}\frac{\partial^{k}}{\partial\beta^{k}}f(\beta){\bigg |}_{\beta=0}=Tr{\bigg (}{\bigg [}r\Gamma(\mathds{1}_d+r\Gamma)^{-1}{\bigg ]}^{k}{\bigg )}
\end{equation}
With a relabeling of the index we get:
\begin{equation}
 p_{m}=\frac{1}{m}\sum_{k=0}^{m-1}{\bigg (}p_{k}\sum_{u=1}^{d}{\bigg [}\frac{r\lambda_{u}}{1+r\lambda_{u}}{\bigg ]}^{m-k}{\bigg )} \ \ \forall \ m\geq1
\end{equation}
which ends the proof.
\end{document}